\documentclass[runningheads]{llncs}

\usepackage[T1]{fontenc}
\usepackage[utf8]{inputenc}
\usepackage{graphicx}
\usepackage{booktabs}
\usepackage{url}
\usepackage{float}
\usepackage[hidelinks]{hyperref}

\begin{document}

\title{Big Enough to Break Out: Tracking the Rising Capability of LLM
Penetration-Testing Agents}

\titlerunning{Tracking the Rising Capability of LLM Pentesting Agents}

\author{Victoria Lovelace \and
Cameron Berryman \and Yuhan You \and Suhas Reddy Adavelly \and Joel Sadler \and Daniel Graham}

\authorrunning{V. Lovelace et al.}

\institute{University of Virginia, Charlottesville VA, USA\\
\email{\{bwg4mg, kqe6rf, mxr9et, xee3yr, wmx9vg, dgg6b\}@virginia.edu}}

\maketitle

\begin{abstract}
Large language model (LLM) agents are increasingly applied to penetration testing,
but we still know little about what they can do or how they fail. We compare two
PentestGPT-based systems: a legacy human-in-the-loop system
running the open-weight Kimi K2.5, and a newer autonomous system running Claude Opus 4.8. 
Across three public targets, the autonomous system solves all three, including the
two the legacy system never finishes. The legacy result is the more surprising of
the two. Even on the machines the legacy system fails to solve, it completes about half the subtasks, while running on ordinary university GPUs with no provider guardrails. 
We can describe the trend but not explain it, since model, harness,
autonomy, and memory architecture all change together. Its direction still points
to the next question: what will limit these agents as they take on more complex tasks? The usual
answer is long-horizon memory, the loss of access to earlier findings during long
attack chains. We test it by adding a coverage-memory layer to both systems, and
neither improves outcomes. In the legacy stalled runs we could review, the limiting factor
appeared to be planning and commitment rather than lost memory: agents held the evidence for a
route forward and never turned it into a concrete exploitation hypothesis, which may
suggest that offensive capability will advance with agents' ability to plan rather
than with better memory. The same subtask scoring that tracks this capability is
available to defenders, who can measure it as it rises instead of waiting to meet
it in the field.

\keywords{Penetration testing \and LLM agents \and Autonomous agents \and
Offensive security \and Open-weight models \and AI safety} \end{abstract}

% =============================================================================
\section{Introduction}
% =============================================================================

Large language models can now automate parts of penetration testing that once required sustained human reasoning. An LLM agent can read tool output, propose commands, revise a plan, and coordinate work across the stages of an assessment. PentestGPT is an influential example. Its original evaluation found that LLMs could handle specific subtasks, such as operating tools and interpreting results. But the models struggled to hold the overall context of an assessment~\cite{deng2024pentestgpt}. Newer systems push for more autonomy through persistent state, automated execution, and updated architectures~\cite{muzsai2024hacksynth,peng2026hackers}.

For a defender, the question is no longer only whether these agents can complete an attack, but how fast that capability is rising and what will limit it next. This paper takes a projection-oriented view: we measure two points on a capability trajectory, then use the gap between them, together with an analysis of the failures that remain, to reason about where the capability is heading.

The first point is a legacy human-in-the-loop PentestGPT harness driven by Kimi K2.5, an openly available model that we ran on ordinary university computing. The second is a newer autonomous harness driven by Claude, a frontier model accessed through a commercial provider. The autonomous configuration completes attack chains the legacy one does not, yet the legacy configuration still works through roughly half the subtasks on machines it never solves. Two features make this progression relevant to future risk. First, that level of capability is already easy to reach, needing only openly available weights and commodity GPUs. Recent work shows that even small models running locally can drive real intrusion behavior~\cite{arat2026}, and the models now in open circulation are considerably larger. Second, the gap points in one direction: better models and harnesses should extend attack depth further.

We are careful about what this comparison can show. No single variable is isolated between the two setups: model, framework, execution, autonomy, and memory all differ, and the targets have public walkthroughs that may appear in training data. We therefore treat the progression as descriptive rather than causal (Section~\ref{sec:threats}), and use it to ask a narrower question: what stops these agents from carrying an assessment through to a more complex objective?

In the original PentestGPT evaluation, loss of session context was the most frequent cause of failed trials~\cite{deng2024pentestgpt}, and more recent work finds that agents can exhaust their context on low-value branches before finishing a chain~\cite{deng2026goodagent}. The autonomous PentestGPT maintainers call a related loss of progress \emph{controller convergence}. If losing access to earlier findings is what bounds depth, then preserving those findings should help. We test that directly by adding a coverage-memory layer to both harnesses. Each keeps a record of the attack surfaces already explored and makes it available during planning, using different integration points because the two frameworks manage memory differently.

This paper makes three contributions.

\begin{enumerate}
\item We measure how far two generations of PentestGPT get on the same three
targets, from a legacy human-in-the-loop system to an autonomous one. The gap
between them gives a concrete starting point for tracking how quickly this
capability is advancing.
\item We design and evaluate a coverage-memory implementation for each
PentestGPT framework. Neither improves outcomes, and the legacy version nearly
triples reasoning time per cycle. We also show why: in the autonomous trials, 
where both conditions ran on every target, the retrieval failure they address never occurs.
\item We extend the original PentestGPT subtask evaluation with repeated-trial
scoring, a transcript-based failure analysis, and exploratory runs on a longer target. The analysis suggests the long-horizon bottleneck may lie in planning rather than memory.
\end{enumerate}

% =============================================================================
\section{Background and Related Work}
% =============================================================================

\subsection{PentestGPT}

PentestGPT is an LLM-based penetration-testing framework, published at USENIX Security 2024~\cite{deng2024pentestgpt}. It automates parts of real-world pentesting workflows. It has three modules. The Reasoning Module maintains a Pentesting Task Tree (PTT), an external record of what the agent has tried, discovered, and should explore next. The Generation Module turns the next task into executable commands with chain-of-thought prompting. The Parsing Module compresses verbose tool output before returning it to the Reasoning Module. The original evaluation covered 13 HackTheBox and VulnHub machines, with 182 subtasks across the OWASP Top 10 and 18 CWE items. PentestGPT beat baseline LLM approaches by holding state through the PTT. But it still suffered from context loss, recency bias, and hallucinated commands during long sessions.

\subsection{LLM Wiki}

Karpathy's LLM Wiki is a persistent, wiki-style knowledge base that the LLM builds and maintains itself~\cite{karpathy2024llmwiki}. The model does not re-read raw sources at query time. It reads each source once and adds to a set of linked Markdown pages, then answers from those pages, so knowledge compounds over time. Our legacy coverage-memory implementation adapts this pattern (Section~\ref{sec:legacy-impl}).

% =============================================================================
\section{Systems}\label{sec:systems}
% =============================================================================

We evaluate two PentestGPT harnesses: the legacy human-in-the-loop implementation, and the maintained autonomous framework documented in July 2026. Both use LLMs to plan the assessment, but they divide the work differently among the models, the human operator, and deterministic code.

\subsection{Legacy PentestGPT}

The legacy harness follows the architecture of the original PentestGPT paper~\cite{deng2024pentestgpt}. It organizes an assessment around a Pentesting Task Tree. It uses three model sessions: one for reasoning, one for command generation, and one for output parsing. The reasoning session keeps the overall plan. The generation session turns the selected task into commands. The parsing session interprets the output. A human operator runs each command and returns its output to the system.

In our experiments the human is only an executor. The operator runs the commands PentestGPT proposes and returns the full output, including errors. The operator does not select actions, correct the model, or change the plan. Planning stays with the LLM, even though a person runs the commands.

\subsection{Autonomous PentestGPT}
\label{sec:autonomous}

The autonomous \texttt{pentestgpt\_agent} framework replaces the three-session workflow~\cite{pentestgpt2026repo}. It uses two model roles and a deterministic controller. The Supervisor reads a projection of the run state. It then selects one task or proposes completion. The Executor receives the selected task, runs it with the available tools, and returns a typed result. Each role runs in a fresh provider session, so provider conversation history never serves as persistent memory.

A non-model controller runs the loop. It reads the Supervisor's decision, validates the plan, leases one task, and invokes the Executor. It then validates the returned trace and evidence before committing the new state. The framework defines six task types: \emph{discover}, \emph{enumerate}, \emph{test}, \emph{exploit}, \emph{verify}, and \emph{recover}. Each task lists a target, objective, completion condition, status, and dependencies. The Supervisor may propose at most one new task per decision, and must select that task immediately. This split lets the models propose actions while deterministic code controls scope, dependencies, evidence, retries, completion, and state transitions.

The authoritative record is a SQLite Memory Kernel, which stores run state, tasks, attempts, observations, and transitions. Provider transcripts and model summaries survive only as diagnostic traces. Claude trials also disable Claude Code's built-in memory, so no provider-held state carries across sessions outside the kernel. Evidence must trace back to real output: an observation has to be an exact, contiguous slice of real command or tool output. The model proposes; the controller decides what enters the record.

\subsection{Controller Convergence}
\label{sec:convergence}

The Memory Kernel keeps the full history, but the Supervisor does not see all of it at each step. Its projection holds the open working set, four recently closed tasks, the required dependency context, six recent observations plus any required basis observations, aggregate history, four recent history items, and four recent diagnostics. The Executor sees even less: its task, that task's basis, up to two observations from earlier attempts on it, and one retry diagnostic~\cite{pentestgpt2026repo}.

This bounded projection is a retrieval limit, not a storage limit. An early observation can stay in SQLite yet leave the Supervisor's view. The documentation calls the resulting loss of progress \emph{controller convergence}. Its symptom is repeated discovery or enumeration of surfaces already examined, instead of progress toward exploitation. The documentation gives two explanations. The first blames limited retrieval. Older coverage leaves the projection, so the planner repeats mapping work it can no longer see. The proposed remedy is to preserve compact coverage information~\cite{pentestgpt2026repo}. The second blames planning. In the maintainers' HTB Enigma run, which we repeat ourselves in Section~\ref{sec:enigma}, the Supervisor kept doing discovery and enumeration but never selected an exploit task. The postmortem calls this over-decomposition, not a memory-capacity failure~\cite{pentestgpt2026enigma}. Both agree that SQLite capacity is not the problem. We carry both forward as competing hypotheses. Our live trials do not settle controller convergence, but they let us ask which hypothesis the evidence favors.

% =============================================================================
\section{The Coverage-Memory Intervention}\label{sec:coverage}
% =============================================================================

We added a coverage-memory layer to both frameworks. Its purpose is to test one idea: that persistent access to earlier findings reduces repeated work. Both versions keep the attack surfaces already explored and make them available during planning. The frameworks manage state differently, so the versions use different integration points and are evaluated separately. We documented the design internally on June 24, 2026, before the maintainers documented a similar direction. The two implementations were developed independently~\cite{pentestgpt2026repo}.

\subsection{Design}

Both implementations have two parts. Storage records what the agent has already examined, as plain Markdown files. Recall makes those notes available to the planner before it picks its next action. The notes last one trial and start empty each run. The store is advisory. It never counts as evidence, satisfies a dependency, or marks the objective complete. Each trial can turn the store on or off, so baseline and coverage-memory conditions run in the same framework.

\subsection{Autonomous Implementation}
\label{sec:auto-impl}

Section~\ref{sec:convergence} explained why early coverage can leave the Supervisor's view even though the kernel keeps it. The autonomous layer restores access to that dropped coverage without touching the canonical record. Writing is automatic: as each observation is committed to the kernel, plain code copies it into the vault, with no model call and no duplicates. Reading is a deliberate Supervisor action. Before proposing another discovery or enumeration task, the Supervisor can consult the vault to check whether a surface was already examined. The Executor cannot, because its job is to carry out the chosen task rather than steer the assessment (Section~\ref{sec:framework-issues}). Because writing is code, the only added model cost is the optional recall (measured in Section~\ref{sec:results}). With the intervention off, there are no vault writes and no recall instruction.

\subsection{Legacy Implementation}
\label{sec:legacy-impl}

In the legacy framework we test the same idea with an agent-maintained wiki. It adapts Karpathy's LLM Wiki pattern~\cite{karpathy2024llmwiki}. The wiki lives inside the reasoning session, not as a separate agent. Before the reasoning session reads the latest tool output, a manager module adds the stored summary of confirmed findings and dead ends to the front of that input. After the session updates the Pentesting Task Tree, but before it picks the next task, a second call asks the model to name any new findings or dead ends. The manager stores that response in the wiki. The wiki starts empty each trial and can be turned on or off with a flag.

This differs from the autonomous version in one way. Reading the wiki is deterministic. But writing new entries takes an extra reasoning-model call on every cycle. So the legacy condition changes two things at once: what the session can see, and how much model work happens per cycle. Its comparison against baseline measures the whole wiki, including that added cost. We report the time and token effects separately in Section~\ref{sec:results}.

% =============================================================================
\section{Methodology}
% =============================================================================

We evaluate two versions of PentestGPT: the legacy human-in-the-loop system running Kimi K2.5, and the newer autonomous system running Claude Opus 4.8 (Section~\ref{sec:autonomous}). Within each version, we compare trials with and without the coverage-memory intervention. Because the versions differ in both model and system design, comparisons between them describe changes in capability rather than their cause. We measure overall success, partial progress, and efficiency from saved trial data and transcript review.

\subsection{Execution Setups}

Section~\ref{sec:systems} describes the two harnesses. Here we give only the experimental details. We accessed Kimi K2.5 through the University of Virginia Research Computing GenAI service. Commands ran from a Kali Linux virtual machine. For steps that needed a graphical interface, the operator described what appeared on screen. Autonomous trials use Claude Opus 4.8 with reasoning effort xhigh. Each Supervisor and Executor call starts in a fresh session, with Claude's built-in memory disabled. Each version is tested under a baseline and a coverage-memory condition. We implement these as in Section~\ref{sec:coverage} and analyze them separately.

Both versions follow the same tool rule. The agent may not use an end-to-end vulnerability scanner, such as OpenVAS or Nessus, that runs most of the assessment on its own. It may use a targeted tool after it finds a specific attack surface. For example, it may run \texttt{sqlmap} against an injection point it found, or a public exploit for a named CVE. Tools are used only when the agent asks for them.

\subsection{Benchmark Targets}

The evaluation uses three public vulnerable machines from prior agent evaluations: Metasploitable 2, Bob, and Tr0ll. They differ in attack structure and difficulty. Bob and Tr0ll have clear attack chains. We score them with fixed, target-specific subtask checklists. Metasploitable 2 has several independent vulnerable services and no single intended attack chain, so it gets no checklist. Every condition solves it, so we use it to compare efficiency with the final outcome held constant. All three have public walkthroughs that may well sit in the training data, so a solved machine is not by itself proof that the agent reasoned its way there. Subtask scoring and transcript review show how far each agent progressed, but neither can rule out that risk (Section~\ref{sec:threats}). A fourth target, HackTheBox Enigma, is used only for the exploratory runs of Section~\ref{sec:enigma} and is not scored against a checklist.

\subsection{Trial Protocol}

Each trial starts with a fixed prompt. The prompt gives the machine name, operating system, IP address, and objective, but no attack plan. The proof of success depends on the target. It requires verified root access, the contents of a named proof file, or direct evidence of remote code execution. Metasploitable 2 uses evidence such as \texttt{id} or \texttt{uname -a} output, because it has no single route or flag. We compare baseline and coverage-memory trials only when they share the target, PentestGPT version, and controller settings. Autonomous sample sizes are reported as collected.

Controller settings are held constant within each target. Autonomous trials use a turn/decision/attempt setting of 4/30/2 for Metasploitable 2 and Bob, and 4/40/2 for Tr0ll, with Claude effort xhigh. The exploratory Enigma runs used a six-turn Supervisor budget, a 60-decision cap, four attempts per task, and a different model (Claude Opus~5), and are reported separately for that reason. The legacy conversation is trimmed to about forty messages, after which earlier content may leave the session. In the legacy setup, a command is stopped after 30 minutes. It may be stopped earlier only when its own output shows a sustained failing pattern, such as repeated connection errors or steadily declining performance. The operator returns all output collected and explains why the command was ended. This choice rests only on the tool's output, not on whether the strategy seems promising. Autonomous trials are bounded by the framework's decision and attempt limits, not by a wall-clock timeout.

\subsection{Metrics and Scoring}

We combine measurements from saved run data with structured review of the full transcripts. The main measures are trial outcome and target-specific subtask completion. When two conditions share an outcome, we compare efficiency. We use reasoning cycles or Supervisor decisions, reasoning time, token use, cost, and wall-clock time. Repeated discovery and failure modes describe agent behavior. They are not proof that controller convergence occurred.

\paragraph{Scripted Metrics.}
For legacy trials, a script records reasoning cycles, commands executed, model reasoning time, and token use. Reasoning time starts when command output returns to the model and ends when the model finishes reasoning about it. It excludes execution and output-return time, and cycles that cross a save-and-resume point. Token counts use the Kimi tokenizer, marked exact or approximate. For autonomous trials, measurements come from the saved run state and provider logs: decisions, task types, token use, cost, and wall-clock time. A separate script measures repeated discovery, reporting a strict count of tasks repeated with the same objective and a loose count of every return to a surface already examined. We review the cases in between by hand.

\paragraph{Subtask and Transcript Scoring.}
We score Bob and Tr0ll with fixed, target-specific subtask checklists. We prepare each checklist before reviewing transcripts, from at least two independent walkthroughs. A subtask gets credit only when the transcript shows command output or other direct evidence of completion. The model's own claim is not enough. We accept alternate valid routes, so scoring measures how far the agent got without forcing one exact path. This follows the original PentestGPT subtask approach. We extend it by scoring individual repeated trials and reviewing failure patterns across both systems. Transcript review also assigns trial outcomes and failure modes. Each judgment is linked to its supporting transcript segment.

\paragraph{Outcomes and Failure Modes.}
Each admitted trial gets one of four outcomes. A trial is a \emph{success} when the agent completes the stated objective and gives direct evidence, such as verified privilege output, a root shell, or the requested proof file. A \emph{stall} is when a trial ends without success after the agent enters a sustained period of no progress.\footnote{We use \emph{stall} for the outcome the draft previously called ``deadlock.'' No progress means the agent produces no new finding and does not try a meaningfully different approach. Changing only a username, path, wordlist, or command option does not count as a different approach. We assign a stall by reviewing the terminal part of the transcript, not by a fixed number of cycles or decisions.} A \emph{generation failure} is when the system produces no actionable task and does not recover after one retry. An \emph{aborted} trial cannot continue because of an outside problem, such as target failure, network loss, provider interruption, or a framework crash. Aborted trials fall under the admissibility rules in Section~\ref{sec:admissibility} and are not counted as agent failures. For every stalled trial, we keep the terminal no-progress sequence with references to the transcript, so the outcome can be reviewed later.

The failure taxonomy adapts categories from the original PentestGPT evaluation. \emph{Context loss}: the agent behaves as though a known result is no longer known. \emph{False command generation}: an invalid or non-executable command. \emph{False output interpretation}: a conclusion the tool output does not support. \emph{Failure to identify the solution path}: relevant findings are present, but the agent does not connect them into a next step. \emph{Hallucinated finding}: the agent asserts an unsupported fact and reasons from it. \emph{Fixation}, or failure to pivot: the agent keeps varying an unsuccessful tactic without changing the broader approach.

\subsection{Trial Admissibility}
\label{sec:admissibility}

We admit a trial as a measurement of agent reasoning only under two conditions: the target works as intended, and the run ends because of the agent's own behavior. We exclude other trials from the reasoning-capability statistics. Infrastructure failures include a required service not starting, a host exposing none of the expected ports, a broken route, or a misconfigured VM. Framework and provider failures include a crash, a turn or decision budget that ends a still-progressing run, a provider usage limit, or a policy refusal. We exclude these as system-level interruptions rather than reasoning failures. Some early autonomous trials were interrupted by provider restrictions on high-risk cybersecurity activity, before we obtained approved access through Anthropic's Cyber Verification Program~\cite{anthropic2026glasswing}. We record these pre-approval refusals as provider failures, not as unsuccessful reasoning trials.

% =============================================================================
\section{Results}
\label{sec:results}
% =============================================================================

Two patterns dominate the results. First, the Kimi legacy system and the Claude autonomous system differ sharply in observed capability. The systems differ in too many ways to say what caused the gap. Second, neither coverage-memory implementation improved outcomes on the targets tested, and both added cost. We also report an exploratory analysis of the unsuccessful legacy trials and of three autonomous runs on a longer target.

\subsection{Capability Progression Across Generations}

The legacy and autonomous systems show a large capability gap on the machines where both ran, though the comparison is confounded. Under the legacy baseline, the Kimi agent solved Metasploitable 2 in all five trials but stalled on Bob and Tr0ll. Under the autonomous baseline, Claude solved all three admitted trials on each of Metasploitable 2, Bob, and Tr0ll. It reached the stated objective in every run.

\begin{figure}[t]
\centering
\includegraphics[width=0.88\textwidth]{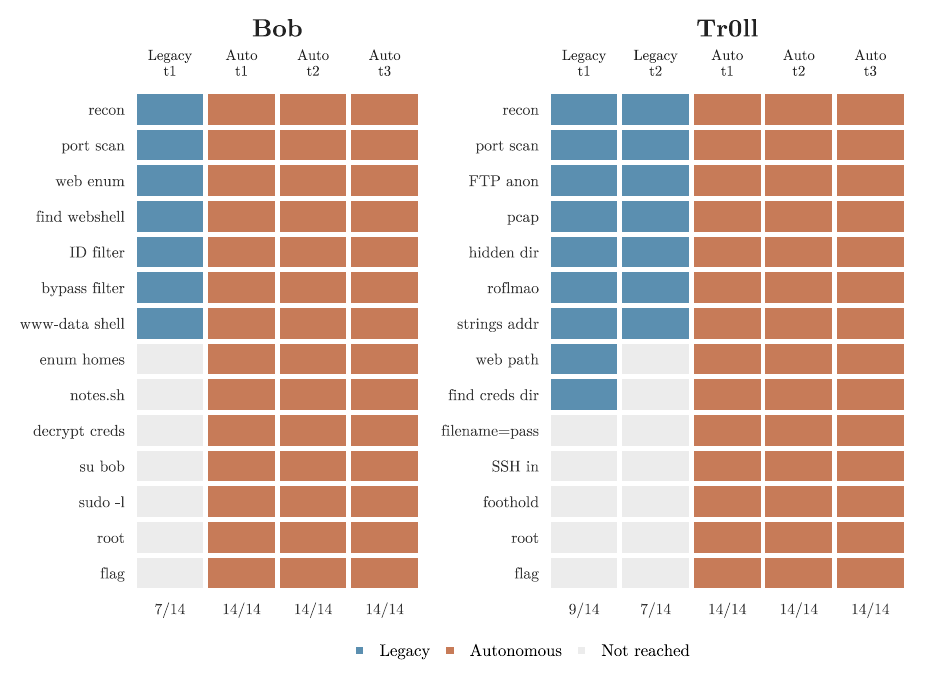}
\caption{Subtask completion by trial on Bob and Tr0ll. Rows are the fixed 14-step checklist in chain order. A cell is filled where the transcript shows direct evidence of completion. Solved autonomous runs are 14/14 by definition.}
\label{fig:grid}
\end{figure}

The subtask counts show the gap even before the final outcome. Kimi reached 7 of 14 subtasks on Bob. It reached 9 and 7 of 14 subtasks across the two Tr0ll trials before stalling. Claude completed all 14 subtasks in every admitted baseline Bob and Tr0ll run. Claude was also much faster, and the margin is conservative. Legacy reasoning time alone, which excludes the time the human spent running commands, still exceeded the full wall-clock time of the autonomous runs. Figure~\ref{fig:grid} shows subtask completion by trial. We evaluate the coverage-memory trials separately in Section~\ref{sec:coverage-results}.

This comparison cannot isolate the cause of the gain, because the setups differ on every axis (Section~\ref{sec:threats}). The directional point survives that confound: on the machines it cannot solve, an accessible open-weight configuration still reaches about half the subtasks, and the newer autonomous configuration completes every chain.

\subsection{Coverage Memory Did Not Improve Outcomes}
\label{sec:coverage-results}

The legacy and autonomous interventions differ in implementation and in added cost. So we evaluate their results separately.

\paragraph{Autonomous Setup.}
Metasploitable 2 is the simplest target. Three baseline and two coverage-memory trials each solved it in exactly three Supervisor decisions. The average cost was \$0.82 for baseline and \$0.79 with coverage. Wall-clock times varied a lot: 2.2 to 11.5 minutes for baseline, and 3.6 to 19.0 with coverage. But the decision counts were identical across the two conditions.

The layer is not entirely free. Its write path adds no model calls, but recall makes the Supervisor query the vault. In the Metasploitable 2 coverage condition, Supervisor output rose by roughly 18\% per decision. Cached Supervisor input roughly doubled, because the prompt now held the retrieved vault content. Total token use stayed near baseline, because the Executor used fewer tokens. So the intervention shifted model use toward the Supervisor rather than cutting total work. Much of the added input was cached, so the dollar effect was smaller still.

On Bob, the intervention also gave no benefit. All six trials reached root from the same \texttt{dev\_shell.php} web-shell foothold as \texttt{www-data}, but the routes from there differed. One run compressed the whole escalation into a single exploit task; another attempted a \texttt{sudo}-via-\texttt{apache2} vector, refuted it, and reached root through a recovered credential over SSH. Table~\ref{tab:results} shows the two conditions close on both decisions and cost. But the means hide a wide spread. Baseline cost ranged from \$1.84 to \$10.28 despite the shared foothold. Three trials per condition are too few to credit the small differences to the intervention rather than to run-to-run variation.

Across the three Bob coverage-memory runs, the Supervisor read at least one coverage page in 5 of 10, 5 of 9, and 6 of 8 of its committed decisions, counted from the recorded tool-invocation logs. The layer was therefore available and actively used.

Tr0ll is the one condition whose numbers favor the intervention. Both conditions solved all three trials and completed all 14 subtasks, so the outcome is identical, but the coverage condition averaged 6.3 Supervisor decisions against 9.0 for baseline, at \$3.04 against \$3.60. Taken alone, a 30\% reduction in decisions looks like the effect the layer was built to produce. We do not read it that way. The mechanism the layer targets never fired: repeated discovery was zero on Tr0ll in both conditions, so there was no repeated mapping work for recall to remove. Against the Bob spread, a 2.7-decision mean difference over three trials is not separable from noise. We therefore report Tr0ll as the case most worth re-testing, not as evidence that coverage recall works.

Across all three targets, repeated discovery was zero under the strict identical-objective measure in both conditions. The looser measure flagged at most four returns per trial, each with new evidence recorded in between. No target produced the behavior the layer was designed to prevent. This is the central limitation of the autonomous experiment: the precondition for the intervention to help never occurred, so these trials are not a live test of controller convergence. They show that coverage recall does not improve outcomes on chains this short. They cannot show whether it would help once earlier coverage actually leaves the projection.

\paragraph{Legacy Setup.}
We ran the legacy coverage-memory condition on Metasploitable 2 only; our work moved to the autonomous framework before the remaining legacy targets were run. On that target the wiki shows a different cost pattern from the autonomous layer. It did not change the mean cycle count, which was 2.4 cycles under both conditions. But it raised mean reasoning time per cycle by 2.9$\times$, from 99.3 to 286.6 seconds. Every wiki trial was slower than every baseline trial across all five trials per condition. Token counts stayed similar, with means of 9,670 versus 9,102 tokens, even though the wiki condition added model calls to update the record. So the cost appeared in reasoning time, not in saved tokens. The legacy wiki thus gave no gain in success or cycle count and added substantial reasoning time. This cost follows from its design. The autonomous layer copies observations into the vault with deterministic code. The legacy wiki instead uses an extra reasoning-model call to generate updates every cycle.

\begin{table}[t]
\centering
\caption{Average results for the (L)egacy and autonomous/(M)odern versions on Bob
and Tr0ll. Metasploitable 2 has no subtask checklist and is discussed in prose
(Section~\ref{sec:coverage-results}).}
\label{tab:results}
\begin{tabular}{llccccccc}
\toprule
Machine & Ver. & Wiki & $n$ & Solved & Time (min)$^{1}$ & Steps$^{2}$ &
Subtasks$^{3}$ & Cost$^{4}$ \\
\midrule
Bob   & L & N & 1 & 0/1 & 155.8 r & 32.0 & 7/14   & n/a \\
Bob   & M & N & 3 & 3/3 & 26.7 w  & 9.3  & 14/14  & \$5.99 \\
Bob   & M & Y & 3 & 3/3 & 21.9 w  & 9.0  & 14/14  & \$5.98 \\
Tr0ll & L & N & 2 & 0/2 & 68.3 r  & 24.5 & 8.0/14 & n/a \\
Tr0ll & M & N & 3 & 3/3 & 21.3 w  & 9.0  & 14/14  & \$3.60 \\
Tr0ll & M & Y & 3 & 3/3 & 19.5 w  & 6.3  & 14/14  & \$3.04 \\
\bottomrule
\end{tabular}

\vspace{0.6em}
\begin{minipage}{\textwidth}
\footnotesize
$^{1}$ r = legacy reasoning time only, a lower bound that excludes human execution time; w = autonomous wall-clock time. $^{2}$ Steps are reasoning cycles for legacy runs and Supervisor decisions for autonomous runs. The two units come from different architectures and are not comparable. $^{3}$ Legacy subtask counts are scored from the run transcripts; solved autonomous runs are 14/14 by definition. $^{4}$ Legacy runs have no dollar cost because Kimi K2.5 ran on university computing rather than a metered provider.
\end{minipage}
\end{table}

\subsection{Failure Patterns in Legacy Stalls}
\label{sec:stalls}

This analysis covers only the legacy system, and not by choice. Every admitted autonomous trial on the benchmark targets succeeded, so the controlled comparison produced no autonomous stalls to examine. Section~\ref{sec:enigma} reports exploratory autonomous runs on a longer target that do stall, but those fall outside the controlled comparison. Claims about planning as a bottleneck therefore rest mainly on the weaker harness.

With that caveat, we reviewed three stalled legacy trials, one on Bob and two on Tr0ll. Fixation appeared in all three, though its form and severity differed.

On Bob, the Kimi agent found a foothold through the \texttt{dev\_shell.php} web shell and identified possible privilege-escalation commands involving \texttt{service apache2} and \texttt{systemctl ssh}. It reached 7 of 14 subtasks before stalling. It then spent roughly six cycles varying the same privilege-escalation direction, with no broader change in strategy. Fixation appeared at least three times. Failure to identify the solution path appeared once: the agent had code execution but did not find the credential chain to root. False output interpretation appeared once: it treated a CVE description as compilable exploit code. The first Tr0ll trial reached 9 of 14 subtasks. It located the directory whose filename is itself the credential, but treated the decoy contents of the file as the password instead, and never recognized the filename. It then ran an SSH brute-force loop for roughly ten cycles (fixation $\geq$3, failure to identify the solution path once).

The second trial reached 7 of 14 and stalled in two distinct phases. Through roughly the first two-thirds of the run it re-enumerated SSH and re-analyzed the packet capture as though the earlier analysis had not happened (context loss $\geq$3, fixation $\geq$2). It then recovered the address embedded in the \texttt{roflmao} binary, but read that address as a memory offset rather than a web path, and spent the remainder of the run tracing it in a debugger; the trial ends mid-disassembly. Its terminal failure is therefore false output interpretation compounded with failure to identify the solution path, not context loss.

In each trial, the transcript held findings relevant to a route forward that the agent did not connect or act on before stalling. The subtask counts are scored from the transcripts and the failure-mode counts are lower bounds. The stall outcomes themselves are confirmed. The sample is small, and the comparable autonomous failures are exploratory (Section~\ref{sec:enigma}), so we cannot draw a broad conclusion. But on these runs, the binding constraint was how the agent selected and committed to a hypothesis, not whether earlier evidence was retained.

\subsection{Framework Issues Identified During Validation}
\label{sec:framework-issues}

We found and fixed two defects in the autonomous framework while validating the coverage layer. The first was a negation-blind task-kind guard. The framework rejected tasks whose objectives disclaimed an action, such as ``do not read the flag,'' because it read the reference to the action as a request to perform it. The second concerned the recall skill, which was initially available in both role workspaces. The Executor sometimes consulted it mid-task, spending turns on vault lookups instead of finishing the work it had been given. In one privilege-escalation task with a tight turn budget, those lookups left it one step short of root. We therefore restricted recall to the Supervisor. This is the final design described in Section~\ref{sec:auto-impl}.

\subsection{Over-Decomposition on a Long Target}
\label{sec:enigma}

The three benchmark targets are short enough that the autonomous planner never stalled. To see whether it fails differently on a longer chain, we ran three exploratory trials against HackTheBox Enigma, a multi-service target with a substantially longer dependency chain, using Claude Opus~5 at effort \texttt{xhigh} with a six-turn Supervisor budget, a 60-decision cap, four attempts per task, and the coverage layer disabled. These are not part of the controlled comparison: the model differs from the rest of the autonomous arm, and the target is an active HackTheBox machine. A run at this decision cap exceeds a single provider usage window, so runs were resumed from the kernel across windows; because state lives in SQLite and every episode starts a fresh session, a resume is not a break in the record. One run was additionally resumed on a
fresh instance of the target after its host crashed, and its rollback removed an exploitation task the pre-crash trajectory had selected. We report these as observations rather than measurements.

All three exhausted the decision budget without reaching root, and all three produced task profiles dominated by breadth: one discovery task, 24 to 26 enumeration tasks, and 27 to 32 test tasks, out of 56 to 59 tasks total. Two runs' final records contain no exploitation task at all, cycling enumeration and testing to the cap. The third selected four, and none produced a foothold: across all three runs no observation records a shell, a privilege check, or either proof file. Repeated discovery was zero in all three runs under the strict identical-objective measure, so the planner never re-ran work it had lost sight of. The looser measure is high, 20 to 25 returns per run to the two surfaces the target exposes, but no two of those tasks share an objective and each follows new recorded evidence. The planner was subdividing the same surfaces, not forgetting them. This is the over-decomposition pattern the maintainers report~\cite{pentestgpt2026enigma}, here in our own runs, and it points at planning rather than at loss of stored findings.

% =============================================================================
\section{Discussion}
% =============================================================================

The coverage-memory layer targets a plausible retrieval failure: as runs grow longer, earlier findings leave the planner's visible context and the agent repeats discovery. Neither implementation improved outcomes, and as Section~\ref{sec:coverage-results} showed, the reason matters more than the result. The failure the layer exists to prevent never occurred in the autonomous trials of the controlled comparison, so there was nothing for it to fix. Two lessons follow. The first is methodological. An intervention aimed at long-horizon memory can only be evaluated on targets long enough for the memory failure to occur, and target length should be chosen by that criterion rather than by nominal difficulty. Every target where we could run both conditions failed this test, which we could establish only after running them. The second is practical. Persistent memory is often treated as a safe default in agent design, but it is not free: the legacy wiki cost 2.9$\times$ the reasoning time per cycle, and the autonomous recall step raised Supervisor output by roughly 18\% per decision. Where retrieval is not actually failing, that is cost without return. Adding memory should follow evidence that the planner is losing information, not precede it.

The stalls reviewed in Section~\ref{sec:stalls} point somewhere else. What limited those runs was not what the agent retained but what it did with it: the information for a route forward was present, and the agent kept decomposing, re-enumerating, or varying a tactic instead of acting on it. Three legacy stalls cannot establish weak commitment as the general cause of controller convergence, but they do point away from memory capacity as the binding constraint, and the exploratory autonomous runs of Section~\ref{sec:enigma} show the same thing in a harness where stored findings were never lost.

This matters for projection. If broader evaluation confirms the pattern, future systems may need to rank hypotheses, commit to an exploitation path, and pivot when evidence contradicts it. Larger or more accessible memory would not be the fix. A planning-focused mechanism changes how the agent chooses among available findings. Coverage recall only changes which findings stay accessible. If depth is bounded by planning rather than memory, the capability trajectory will track models' planning ability, not memory engineering. The present data support only a descriptive progression.

There is a defensive reading of the same result. The configuration that needs no provider approval is also the one a defender can stand up in a lab. Running it against a known machine shows in a few hours where a current agent stops and why it stops there. Our transcripts put that point at hypothesis selection rather than at reconnaissance or at memory, which is more useful to a defender than the final score. The same checklists and failure categories can be rerun on each new model, so a curriculum built this way tracks the tools instead of going stale alongside them.

% =============================================================================
\section{Threats to Validity}
\label{sec:threats}
% =============================================================================

\paragraph{Benchmark contamination.}
Training data may have held the exact solutions to the evaluated machines~\cite{chen2025contamination}. These machines are well documented, especially Metasploitable 2, and have public walkthroughs. If those appeared in Kimi's or Claude's training data, the models may have reproduced parts of a known solution rather than deriving each step from evidence. No public target can remove this threat. Subtask checklists and transcript review show how far each agent got, but they cannot tell us whether a successful action was reasoned or recalled.

\paragraph{Small samples and no statistical inference.}
Every condition here has between one and five trials, including a single admitted trial for legacy Bob. We therefore report means and ranges and draw no statistical inferences. The outcome results are the most robust, being categorical and unanimous within each condition; the efficiency comparisons are the least, and we treat apparent differences in decisions or cost as provisional.

\paragraph{Confounded cross-setup comparison.}
The two setups differ on every axis available: model, execution paradigm, framework, autonomy, and memory system. Nothing varies alone, so no single factor can be named as the cause of the capability difference. Our open-weight model runs only in the weaker harness, and our frontier model only in the stronger one. So these results cannot establish how capable open-weight models are in general, nor attribute the gap to the models rather than the harnesses. What they do show is that one accessible, unguardrailed configuration solves Metasploitable 2 outright and reaches about half the subtasks on the two machines it fails. We treat the comparison as a descriptive progression, not a causal result.

\paragraph{Non-identical controller limits.}
Autonomous controller settings varied across machines: turn/decision/attempt of 4/30/2 on Metasploitable 2 and Bob, 4/40/2 on Tr0ll, and a six-turn, 60-decision, four-attempt setting with a different model (Claude Opus~5) for the exploratory Enigma runs. Within the benchmark set they were constant across conditions on the same target, so within-target comparisons are valid; cross-target comparisons are not. Turn limits are also imperfect bounds, since an agent may take several tool actions per turn.

\paragraph{Scoring judgment.}
Subtask and failure-mode scoring needs some human judgment. Each subtask requires direct evidence, but deciding whether two actions share a broader approach, or whether a finding supports a solution path, is less objective. We keep the supporting transcript segments so these calls stay reviewable. Independent scoring would strengthen future evaluations.

% =============================================================================
\section{Future Work}
\label{sec:future}
% =============================================================================

The most pressing need is a custom target with no public walkthrough. Without one, the capability progression cannot rule out memorized solutions. Running two or more models through a single autonomous framework would then separate the causes we cannot currently untangle, holding framework, execution, and memory constant while varying only the model.

The retrieval hypothesis needs a longer target. Early findings have to be able to leave the Supervisor's projection before the hypothesis can be tested at all. The three benchmark machines are too short, and repeated discovery is zero in our autonomous trials. Enigma is long enough. All three runs in Section~\ref{sec:enigma} stalled, but we ran them at baseline, with the coverage layer off. Running that same target with the layer on is the obvious next experiment. It varies one condition on a machine we have already characterized.

Finally, the legacy stalls suggest a planning-focused intervention, since fixation appeared in all three reviewed trials. This evidence is limited. So broader transcript analysis should first check whether weak hypothesis selection and fixation recur across more models, targets, and runs. If they do, future work could test a mechanism that keeps a concrete exploitation hypothesis, caps repeated effort on an unproductive approach, and forces a pivot when that budget runs out.

% =============================================================================
\section{Conclusion}
% =============================================================================

We tracked two points on the capability trajectory of LLM penetration-testing agents. The autonomous frontier-model configuration finished every attack chain, including the two the legacy one never did. The legacy configuration is the one that should concern a defender, not because it performed better, but because nothing gates it. An openly available model with nothing more than commodity university GPUs worked through roughly half of the subtasks on the boxes it failed to solve, and no provider was present to refuse the requests. We cannot say whether the model or the harness accounts for the gap. What we can say is that halfway is where the openly available configuration sits today, and that these systems are not getting worse.

We then tested the usual explanation for where these agents stop, long-horizon memory, and did not confirm it. Adding a coverage-memory layer improved neither harness, and in the legacy system it nearly tripled the reasoning time each cycle took. In the autonomous runs no finding ever left the planner's view, so recall had nothing to restore. Where we could inspect failures directly, they looked less like forgetting than like a failure to commit, with the evidence for a route forward sitting unused in the transcript. If that holds more broadly, capability will advance with models' ability to plan rather than with better memory engineering, and measuring it will require targets no model has already read the answer to.

% =============================================================================

\end{document}